\documentclass[prd,preprint,nofootinbib]{revtex4}
\usepackage{graphicx}
\usepackage{amssymb}

\begin{document}
\begin{flushright}
\hfill DESY 06-212 
\end{flushright}

\title{Constraint on the Effective Number of Neutrino Species from the WMAP and SDSS LRG Power Spectra}

\author{Kazuhide Ichikawa, Masahiro Kawasaki}
\affiliation{
Institute for Cosmic Ray Research, University of Tokyo,
Kashiwa 277 8582, Japan}
\author{Fuminobu Takahashi}
\affiliation{Deutsches Elektronen Synchrotron DESY, Notkestrasse 85,
22607 Hamburg, Germany}

\date{\today}

\vskip10mm
\begin{abstract}
We derive constraint on the effective number of neutrino species
$N_\nu$ from the cosmic microwave background power spectrum of the
WMAP and galaxy clustering power spectrum of the SDSS luminous red
galaxies (LRGs). Using these two latest data sets of CMB and galaxy
clustering alone, we obtain the limit $0.9 < N_\nu < 8.2$ (95\% C.L.)
for the power-law $\Lambda$CDM flat universe, with no external
prior. The lower limit corresponds to the lower bound on the reheating
temperature of the universe $T_R > 2$ MeV.
\end{abstract}
\maketitle


\section{Introduction}

The standard model of cosmology with the concordance set of parameters
can successfully reproduce a broad range of the cosmological data such
as the big bang nucleosynthesis (BBN), the cosmic microwave background
(CMB) anisotropies and large scale structure (LSS).  The relativistic
degrees of freedom present after the BBN epoch in the standard
cosmology are photons and three generations of neutrinos. In models of
particle physics/cosmology, however, there are many candidates that
could additionally contribute to the relativistic components of the
universe: sterile neutrinos~\cite{sterile}, gravitational waves,
(pseudo-)Nambu-Goldstone bosons such as axions~\cite{Peccei:1977hh}
and majorons~\cite{majoron}, and the active neutrinos themselves if
they have large lepton asymmetries~\cite{lepton-asym}.  Furthermore,
the energy density of the relativistic particles can be smaller than
in the standard cosmology, if the thermalization of the neutrinos are
ineffective as in the MeV-scale reheating
scenarios~\cite{Kawasaki:1999na, Ichikawa:2005vw}.  In fact, in a
certain class of models, especially those accompanied by the late-time
entropy production~\cite{Lyth:1995ka}, the (final) reheating
temperature tends to be quite low, and it often falls in the MeV
range.  Therefore it is of great importance to study the possible
effects of varying the effective number of the relativistic particles
on the cosmological observations, not only to make the standard
cosmology more established, but also to probe and constrain a certain
class of models in the particle physics/cosmology.

Recent precise observations of the CMB anisotropies and LSS make it
possible to measure the relativistic degree of freedom in the universe
through its effects on the growth of cosmological perturbations. 
These effects come from the fact that the density perturbation does not 
grow (the gravitational potential decays) during the radiation-dominated era. 
Specifically, more relativistic degree of freedom causes more 
early integrated Sachs-Wolfe effect on the CMB power spectrum,
which leads to higher first peak height.
Also, since it delays the epoch of the matter-radiation equality and makes the horizon
 at that time larger, the turnover position of the matter power
spectrum is shifted to larger scales and the power at smaller scales are
suppressed. Therefore, by observing CMB and LSS, we can measure the 
relativistic degree of freedom during the structure formation. In detail, 
assuming the smallest scale relevant to our observations to be about 5\,Mpc,
since the structure formation of that scale begins around the temperature $T\approx 20$\,eV
 (at which the scale enters the horizon), these observations probe the relativistic
 degree of freedom at $T \lesssim 20$\,eV. Thus, CMB and LSS can measure the relativistic 
 degree of freedom independently of another well-known probe, BBN, which measures 
it in much earlier universe around $T=O({\rm MeV})$.

In this paper, we analyze the most recent data sets of CMB and LSS,
respectively using the Wilkinson Microwave Anisotropy Probe (WMAP)
3-year data
\cite{Spergel:2006hy,Page:2006hz,Hinshaw:2006ia,Jarosik:2006ib} and
the Sloan Digital Sky Survey (SDSS) luminous red galaxies (LRGs) power
spectrum data \cite{Tegmark:2006az}, and would like to discuss the
constraints from them. 
The WMAP data now can give clear features of 
the first and second peaks of the CMB power spectrum. In particular, the precision
around the first peak has been already cosmic variance dominated and so
has been the measurement of the early integrated Sachs-Wolfe effect. 
However, this is not the case for the measurement of the relativistic degree of freedom
since it is almost completely degenerate with the value of the matter density. 
We would like to demonstrate this degeneracy by the WMAP data alone 
analysis. Then, as is well known, since the matter power spectrum has somewhat 
different degeneracy pattern from the CMB, it is broken by combining the CMB with 
the matter power spectrum data. We would like to see how and how much the degeneracy 
is broken by adding the matter power of the SDSS LRG sample.
Related analyses of earlier data sets are found for example in
Refs.~\cite{Kneller:2001cd,Hannestad:2001hn,Bowen:2001in} using
pre-WMAP data, in
Refs.~\cite{Crotty:2003th,Pierpaoli:2003kw,Hannestad:2003xv,Barger:2003zg,
Crotty:2004gm,Hannestad:2005jj} using WMAP 1st year data, and
\cite{Spergel:2006hy,Seljak:2006bg,Hannestad:2006mi,Cirelli:2006kt}
using WMAP 3-year data, with various combinations of astrophysical
data sets and priors for cosmological parameters.

 The new point of our analysis on the relativistic degree of freedom in the universe is 
 to use the power spectrum of the SDSS LRGs \cite{Tegmark:2006az}. 
 Although there are similar analyses using the 
 power spectrum of the SDSS main galaxies \cite{Tegmark:2003uf} and/or the one of 
 the 2dF Galaxy Redshift Survey (2dFGRS) \cite{Cole:2005sx}, it is important to 
 revisit the issue with the new power spectrum data. This is because not only have the 
 LRGs more statistical constraining power (the effective volume of 
 the LRG survey is about 6 times larger than that of the SDSS main galaxy sample and
 over 10 times larger than that of the 2dFGRS \cite{Tegmark:2006az}), but also there
 seems to have been a tension between the power spectra of the 2dFGRS and the SDSS main galaxies
  \cite{Cole:2005sx,Cole:2006kn}. The discrepant measurements of the relativistic degree of freedom 
 between these two samples as found in Refs.~\cite{Spergel:2006hy,Seljak:2006bg} (we 
 show the result of Ref.~\cite{Seljak:2006bg} in Table \ref{tab:Nnucomparison}. Compare the 4th and 5th lines)
 are considered to be caused by this tension in the power spectra. 
 It is shown in Ref.~\cite{Cole:2006kn} that the discrepancy is due to the scale dependent bias
  which was not taken into account in the SDSS main galaxy analysis. 
  The LRG analysis in Ref.~\cite{Tegmark:2006az} models this effect of scale dependent bias in the same
  way as the 2dFGRS analysis and they found the extracted cosmological parameters, especially the matter density,
 are in excellent agreement with those from WMAP alone and WMAP\,+\,2dFGRS. 
 Therefore, it is useful and of great importance to investigate the relativistic degree of freedom using 
 the LRG power spectrum and see whether the discrepancy in its derived value from the two galaxy surveys
is resolved to give a reliable constraint.\footnote{
At present, the scale dependent bias is modeled in a very phenomenological manner and more detailed 
modeling is considered to be required. However, this is the on-going issue in the community and beyond the scope
of our paper.
} 

We describe our analysis method in Sec.~\ref{sec:analysis} and our results are presented in Sec.~\ref{sec:result}.
In Sec.~\ref{sec:mevreh}, we briefly review a cosmological scenario with low (MeV-scale) reheating temperature and
explain how the relativistic degree of freedom is modified compared to the standard case. 
In Sec.~\ref{sec:discussion}, we discuss our result in relation to previous works and see
how much the current observations allow the relativistic component of the universe to deviate from the 
value in the standard cosmology. We also emphasize its implication for the lower bound on the 
reheating temperature. 

\section{Analysis} \label{sec:analysis}

The quantity we try to constrain in this paper is the effective number of neutrino species, $N_\nu$, which is widely used to quantify the energy density of the relativistic component in the early universe. It is given by $N_\nu = (\rho_{\rm rel}-\rho_\gamma)/\rho_{\nu,{\rm thm}}$, where $\rho_\gamma$ is the photon energy density, $\rho_{\rm rel}$ is the total energy density of photons, three active species of neutrinos and extra relativistic contribution, and $\rho_{\nu,{\rm thm}}$ is defined as  $\rho_{\nu,{\rm thm}}=(7\pi^2/120) (4/11)^{4/3} T ^4$ using the photon temperature $T$ after the electron-positron annihilation. $\rho_{\nu,{\rm thm}}$ corresponds to the energy density of a single species of neutrino assuming that neutrinos are completely decoupled from the electromagnetic plasma before the electron-positron annihilation takes place and they obey Fermi-Dirac distribution.

We constrain $N_\nu$ in the flat $\Lambda$CDM universe with the initial perturbation power spectrum which is adiabatic and described by power law. This model has 6 cosmological parameters, the baryon density $\omega_b$, the matter density $\omega_m$, the normalized Hubble constant $h$, the reionization optical depth $\tau$, the scalar spectral index of primordial perturbation power spectrum $n_s$ and its amplitude $A$ ($\omega = \Omega h^2$, where $\Omega$ is the energy density normalized by the critical density). Theoretical CMB and matter power spectra are calculated by the CMBFAST code \cite{Seljak:1996is} and $\chi^2$ by the likelihood codes of the WMAP 3-year data \cite{Page:2006hz,Hinshaw:2006ia,Jarosik:2006ib} and of the SDSS LRG power spectrum data \cite{Tegmark:2006az}. We apply modeling of non-linearity and scale dependent bias as in Ref.~\cite{Tegmark:2006az} to the linear matter power spectrum before fitting to the LRG data. Since we omit the process of ``dewiggling" (which is found to be justified in Sec.~\ref{sec:result}), this modeling has two parameters, galaxy bias factor $b$ and non-linear correction factor $Q_{\rm nl}$. Specifically, we connect the linear matter power spectrum $P_{\rm lin}(k)$ and the galaxy power spectrum $P_{\rm gal}(k)$ by
\begin{eqnarray}
P_{\rm gal}(k) = b^2\, \frac{1+Q_{\rm nl}\, k^2}{1+ 1.4\, k} P_{\rm lin}(k).
\end{eqnarray}
 We calculate the $\chi^2$ as functions of $N_\nu$ by marginalizing over the above parameters (6 parameters for WMAP alone and 8 for WMAP+SDSS). The marginalization is carried out by the Brent minimization \cite{brent} modified to be applicable to multi-dimension parameter space as described in Ref.~\cite{Ichikawa:2004zi}.

\begin{figure}
\begin{center}
\includegraphics{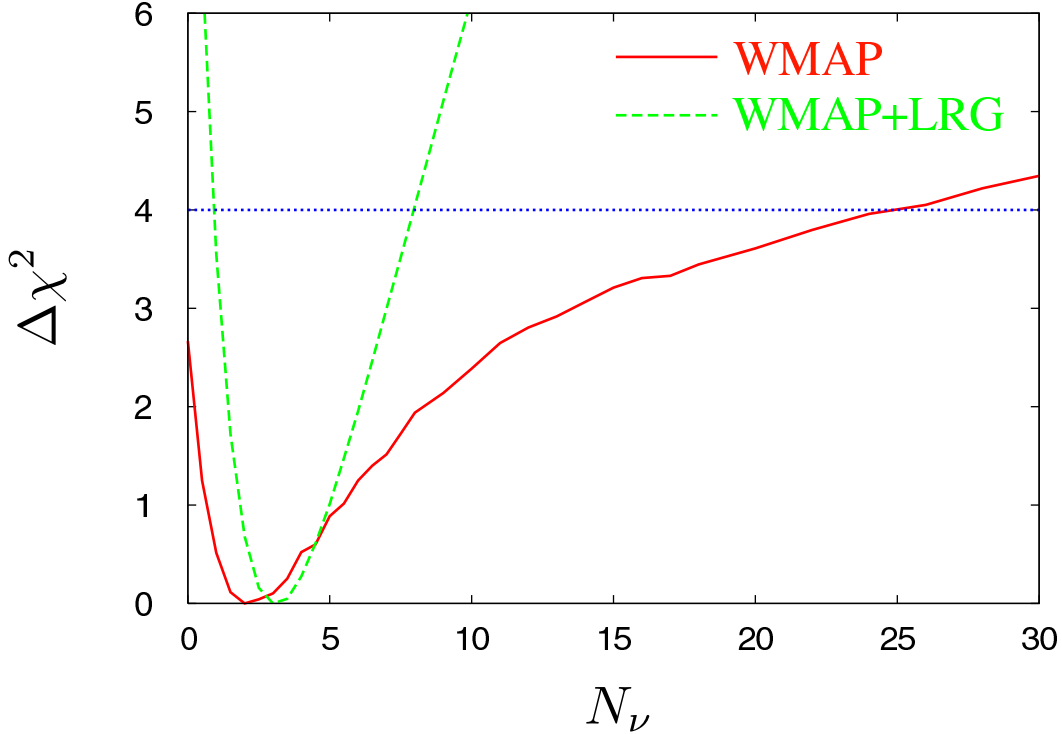} 
\end{center}
\caption{$\Delta \chi^2$ as functions of $N_\nu$. The red solid line uses the WMAP 3-year data alone and the green dashed line uses WMAP 3-year and SDSS LRG power spectrum.}
\label{fig:chi2}
\end{figure}

\section{Result} \label{sec:result}

\begin{figure}
\begin{center}
\includegraphics[width = 12cm]{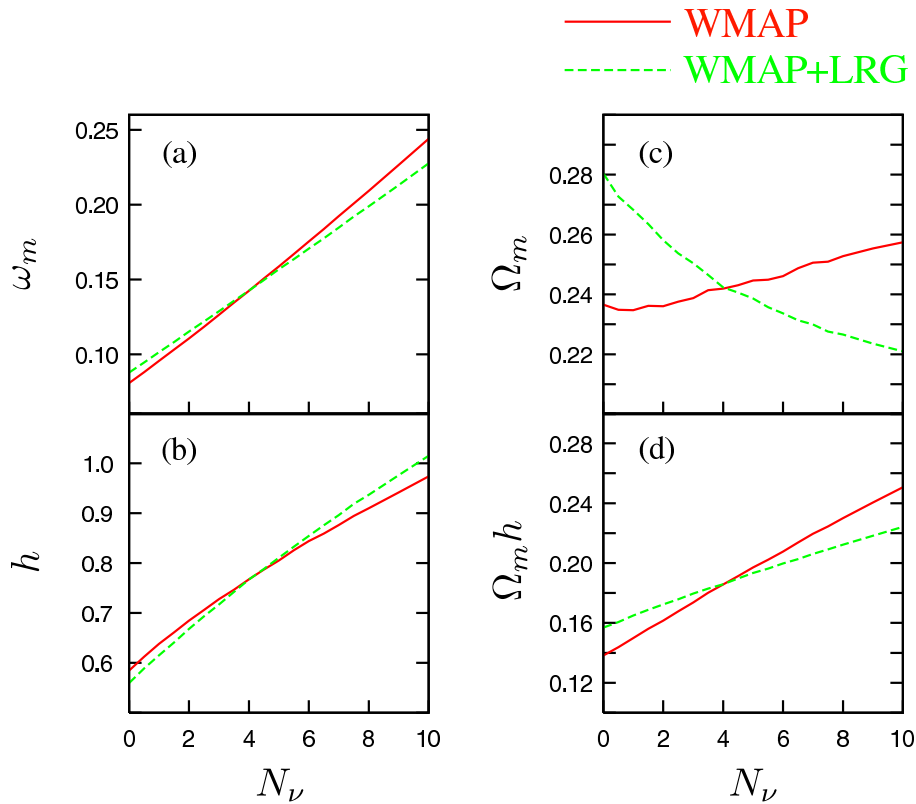} 
\end{center}
\caption{The best fit values of some of the cosmological parameters as functions of $N_\nu$. The red solid lines are for the WMAP 3-year data alone and the green dashed lines are for WMAP 3-year and SDSS LRG power spectrum combined.}
\label{fig:Nnu_parbest}
\end{figure}

We show the results of $\chi^2$ minimization in Fig.~\ref{fig:chi2}. We give the values of some of the best fit cosmological parameters as functions of $N_\nu$ in Fig.~\ref{fig:Nnu_parbest}. We have checked that the results for standard three neutrino species agree with the WMAP \cite{Spergel:2006hy} and SDSS \cite{Tegmark:2006az} groups' analyses. For the WMAP 3-year alone case, it has been checked in Ref.~\cite{Fukugita:2006rm} that the best fit $\chi^2$ and parameters agree. With regard to WMAP and LRG combined analysis, our best fit parameter values for three neutrino species are $\omega_b=0.0222 \pm 0.0007$, $\omega_m=0.1288 \pm 0.0044$, $h=0.718 \pm 0.018$, $\tau = 0.088 \pm 0.029$, $n_s=0.958 \pm 0.016$, $\sigma_8 = 0.770 \pm 0.033$ (we report here $\sigma_8$ instead of $A$ to compare with Ref.~\cite{Tegmark:2006az}), $b=1.877 \pm 0.065$ and $Q_{\rm nl}=30.4 \pm 3.5$. The central values fall well within the 1$\sigma$ ranges of the constraints derived in Ref.~\cite{Tegmark:2006az} and the 1$\sigma$ errors are almost identical to those quoted in Ref.~\cite{Tegmark:2006az}. This empirically shows that the ``dewiggling" we mentioned in Sec.~\ref{sec:analysis} for the non-linear modeling can safely be neglected as is documented in the likelihood code of Ref.~\cite{Tegmark:2006az}. This makes sense as follows. Since the process of the dewiggling mainly decreases the amplitude of the acoustic oscillations in the matter power spectrum, it mostly affects the parameter estimation of $\omega_b$. However, $\omega_b$ is more precisely determined by the CMB when we do the combined analysis so neglecting the dewiggling does not affect much the parameter estimation using the present galaxy clustering data.

The limits corresponding to $\Delta \chi^2 = 4$ are $N_\nu < 25$ for WMAP 3-year alone and $0.8 < N_\nu < 8.0$ for WMAP and SDSS LRG combined. Since $\chi^2$ functions show some asymmetric features, we derive 95\% confidence limits by integrating the likelihood functions ${\cal L} = \exp({-\Delta \chi^2/2})$. This yields 95\% C.L. bound of $N_\nu < 42$ for WMAP alone and $0.9 < N_\nu < 8.2$ for WMAP+LRG.

We observe that the CMB alone constraint is very weak and the LSS data significantly reduces the allowed region. We note that our WMAP 3-year limit is somewhat weaker than the earlier constraints quoted as CMB alone limit \cite{Hannestad:2001hn,Bowen:2001in,Crotty:2003th,Hannestad:2003xv,Barger:2003zg}, even though they use data before the WMAP 3-year release. We compiled them in Table \ref{tab:cmbalone}. For example, Ref.~\cite{Crotty:2003th} has derived $N_\nu < 9$ (95\% C.L.) using the WMAP 1-year data alone, much more stringent than our bound $N_\nu < 42$. We can ascribe this apparent discrepancy to the prior on $h$ adopted by Ref.~\cite{Crotty:2003th}, $h<0.9$, which affects $N_\nu$ constraint through the well-known $h-N_\nu$ degeneracy. We find this degeneracy in our analysis too as in Fig.~\ref{fig:Nnu_parbest} (b) (although we show the result for $N_\nu \leq 10$, for example, $N_\nu = 22$ can fit the WMAP 3-year data with $h \sim 1.3$). They are positively correlated which is explained as follows. The effect of increasing $N_\nu$ on CMB  power spectrum is cancelled by increasing $\omega_m$ so that the epoch of matter-radiation equality occurs at the same redshift. Then, $h$ has to be increased so that $\Omega_m \sim 0.25$ leading to acoustic peaks at observed positions in the flat universe. These features are explicitly demonstrated by the solid lines in Fig.~\ref{fig:Nnu_parbest} (a)--(c).

The degeneracy is broken by adding LSS information as is clearly seen in Fig.~\ref{fig:chi2}. This can be understood by another well-known fact that the shape of the matter power spectrum is determined by the combination $\Omega_m h$ rather than $\omega_m = \Omega_m h^2$.
When $\omega_m$ is varied, obviously it is impossible to find $h$ which preserves both $\Omega_m$ and $\Omega_m h$.
Thus, when we try to fit the CMB and LSS simultaneously, $h$ faces the dilemma of fitting the CMB ($\Omega_m$) or the LSS ($\Omega_m h$).
We can see this in the panels (c) and (d) of Fig.~\ref{fig:Nnu_parbest}. 

Our final result, the constraint on $N_\nu$ from the WMAP 3-year CMB power spectrum and the SDSS LRG power spectrum, can be summarized as 
\begin{eqnarray}
0.9 < N_\nu < 8.2 \quad {\rm or} \quad N_\nu = 3.1^{+5.1}_{-2.2} \label{eq:Nnu_constraint}
\end{eqnarray}
 at 95\% C.L., whose center value is quite close to the standard model of three active neutrino species. We will be discussing the constraint in connection with other works in Sec.~\ref{sec:discussion}. 

\begin{table}[htdp]
\caption{Summary of CMB alone limits.}
\begin{center}
\begin{tabular}{|c|c|c|c|}
\hline
  & CMB data & 95\% limit & Prior on $h$ \\
  \hline
Hannestad \cite{Hannestad:2001hn}   & pre-WMAP & $N_\nu < 19$ or 24 & $0.4<h<0.9$ \\
\hline
Bowen et al. \cite{Bowen:2001in} & pre-WMAP & $0.04 < N_\nu < 13.37$ & $0.4<h<0.95$, $h=0.65\pm 0.2$ Gaussian \\
  \hline
  Crotty et al. \cite{Crotty:2003th} & WMAP1 & $N_\nu < 9$ & $0.5 < h < 0.9$ \\
  \hline
  Hannestad \cite{Hannestad:2003xv} & WMAP1 & $N_\nu < 8.8$ & $0.5 < h < 0.85$ \\
\hline
  Barger et al. \cite{Barger:2003zg} & WMAP1 & $0.9 < N_\nu < 8.3$ & $0.64 < h < 0.8$ \\
  \hline
  \hline
  This paper & WMAP3 & $N_\nu < 42$ & NONE \\
  \hline
\end{tabular}
\label{tab:cmbalone}
\end{center}
\end{table}

\section{Effective number of neutrinos in the universe with low reheating temperature} \label{sec:mevreh}
Before we move on to discuss our result in the next section, it will be useful to review the low (MeV-scale)
reheating scenario. In this scenario, the effective number of neutrinos $N_\nu$ can deviate from the standard value. 
We would like to briefly explain this scenario and how $N_\nu$ and the reheating temperature $T_R$ are related. 
For more details, we refer to Refs.~\cite{Kawasaki:1999na,Ichikawa:2005vw}.

The standard big bang model assumes that the universe
was once dominated by thermal radiation composed of
photons, electrons, neutrinos, and their antiparticles.
The reheating temperature is the temperature at which the universe becomes such radiation dominated state
and it is usually assumed to be so high that every particle species is in thermal equilibrium. 
In particular, neutrinos are considered to obey Fermi distribution.

What if the reheating temperature is lower, say, several MeV? In
contrast to electrons that are always (at least until the
temperature drops below a few eV) in thermal contact
with photons via electromagnetic forces, neutrinos interact
with electrons and themselves only through the weak interaction.
The decoupling temperature of the neutrinos
should be around 3 MeV for the electron neutrinos and
5 MeV for the muon and tau neutrinos, respectively (the difference comes from the fact that the electron
neutrinos have additional charged current interaction with
electrons). Therefore the neutrinos might not be fully thermalized and lead to $N_\nu < 3$
if the reheating temperature is in the MeV range. 

In fact, the reheating temperature as low as a few MeV can be found in many cosmological scenarios.
To avoid the overproduction of the unwanted relics such as
the gravitinos, one needs to require the reheating temperature low
enough~\footnote{This is the case if the gravitinos are thermally produced~\cite{Weinberg:zq, Krauss:1983ik}.
On the other hand, when the gravitinos are non-thermally produced by inflaton decay,
lower reheating temperature leads to more gravitinos, making the gravitino-overproduction
severer~\cite{Kawasaki:2006gs,Dine:2006ii,Endo:2006tf,Endo:2006qk}.}.  In extreme cases it may be in the MeV range.  Further, the
thermal history of the universe may not be so simple that the universe
might have underwent several stages of the reheating, and the final reheating
temperature may be very low. For instance, late-time entropy production~\cite{Lyth:1995ka} 
is one of the plausible ways to solve problems associated with the unwanted
relics, and the reheating temperature often falls in the MeV scale.

\begin{figure}
\begin{center}
\includegraphics{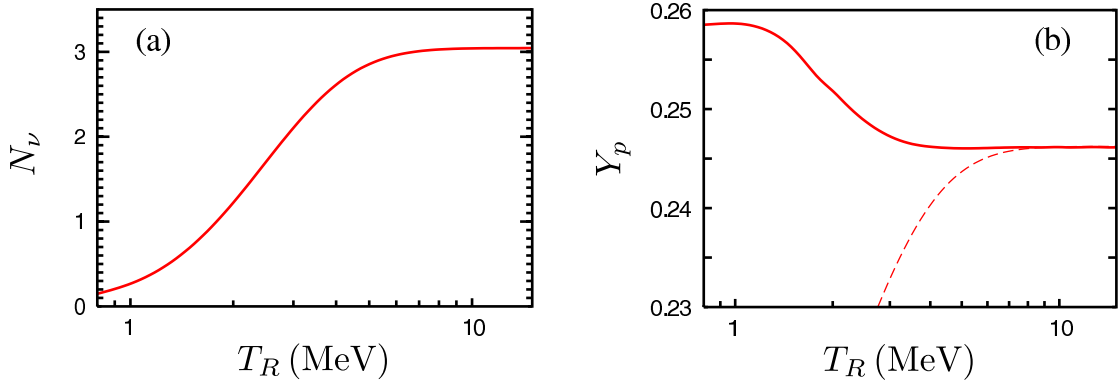} 
\end{center}
\caption{Taken from the calculation in Ref.~\cite{Ichikawa:2005vw}. (a) The relation between the effective neutrino number $N_\nu$ and the reheating temperature $T_R$.  (b) The solid line shows the $^4$He abundance $Y_p$ as a function of the reheating temperature $T_R$. The dashed line is calculated with Fermi distributed neutrinos with $N_\nu$ of the panel (a) (namely, only the change in the expansion rate due to the
incomplete thermalization is taken into account). The baryon-to-photon ratio is fixed at $\eta = 5 \times 10^{-10}$. }
\label{fig:TR_Nnu_He4}
\end{figure}

In Ref.~\cite{Ichikawa:2005vw}, we have calculated how much neutrinos are thermalized 
when $T_R = O({\rm MeV})$ and have derived the relation between $T_R$ and $N_\nu$
which is shown in Fig.~\ref{fig:TR_Nnu_He4} (a).
Specifically, we have solved numerically the momentum dependent
Boltzmann equations for neutrino density matrix, fully
taking account of neutrino oscillations. 
For later convenience, we also show the $^4$He abundance $Y_p$ in the MeV reheating scenario in Fig.~\ref{fig:TR_Nnu_He4} (b).
It should be noted that $Y_p$ increases while $N_\nu$ decreases in this scenario. This is in contrast to 
the conventional non-standard $N_\nu$ scenario where decreasing $N_\nu$ accompanies decreasing $Y_p$. 
The difference occurs as follows. Since the latter assumes the thermal (Fermi) distribution for neutrinos
as in the standard cosmology, only the expansion rate is modified and particularly it has the neutron-proton conversion rate
identical to the standard one. Meanwhile, since the MeV reheating scenario makes the neutrino distribution less thermalized one, 
the neutron-proton conversion rate is significantly modified in addition to the expansion rate.
To elucidate the effect of the modified neutron-proton conversion rate, we draw the dashed line in Fig.~\ref{fig:TR_Nnu_He4} (b)
which expresses (fictitious) $Y_p$ when we include only the change in the expansion rate. 

We can now convert our constraint on $N_\nu$, Eq.~(\ref{eq:Nnu_constraint}), into the lower bound 
on $T_R$ using Fig.~\ref{fig:TR_Nnu_He4} (a):
\begin{eqnarray}
T_R > 2\, {\rm MeV}.
\end{eqnarray}
We will discuss this CMB+LSS constraint on $T_R$, paying particular attention to the comparison
with BBN bound, in the next section.

\section{Discussion}\label{sec:discussion}

We have shown that new data of the SDSS LRG power spectrum can considerably shrink the allowed region of $N_\nu$ from the one obtained using the WMAP 3-year data alone by a factor of six. In terms of the extra relativistic particle species other than three species of active neutrinos, the LRG data reduces the upper limit by a factor $\approx 7.5$, from 39 to 5.2. Moreover, combining with the LRG data gives a finite lower limit on the effective neutrino number, $N_\nu > 0.9$. This translates into the lower bound on the reheating temperature of the universe, $T_R > 2$ MeV as described in Sec.~\ref{sec:mevreh}. 

A comparison to Ref.~\cite{Seljak:2006bg} who has reported the constraints using earlier data sets is in order. They have provided constraints on $N_\nu$ using various combinations of cosmological data sets, which we compiled in Table \ref{tab:Nnucomparison}. We can summarize their finding that their Ly$\alpha$ data \cite{McDonald:2004xn} and/or the galaxy clustering power spectrum data from the SDSS main sample \cite{Tegmark:2003uf} prefer $N_\nu > 3$ at more than 95\% confidence level whereas the 2dF galaxy power spectrum \cite{Cole:2005sx} does not show such a non-standard feature. Our new constraint is quite similar to the latter, WMAP3+2dF(+supernovae) constraint of $N_\nu = 3.2^{+3.6}_{-2.3}$ (95\% C.L.).
 This result is reasonable since the SDSS main galaxy power favors significantly higher value of $\Omega_m$ than the 2dF power \cite{Spergel:2006hy} but the SDSS LRG power gives $\Omega_m$ which is close to the 2dF value \cite{Tegmark:2006az}. The robustness of the estimation of $\Omega_m$ from the SDSS LRG clustering is thoroughly tested by means of the power spectrum shape \cite{Percival:2006gt} and the baryon acoustic oscillations \cite{Percival:2006gs}. Since galaxy clustering basically measures the matter-radiation equality, this robustness is considered to be transferred to our estimation of $N_\nu$. We can conclude that although the constraints from both galaxy surveys has converged with central values around the standard value of three, allowed regions are large enough to cover the constraints obtained with Ly$\alpha$ forest data whose central values are around 5. We have to wait for more study on the Ly$\alpha$ forest analysis and future CMB/LSS observations (the PLANCK sensitivity for $N_\nu$ is forecasted to be 0.2, see e.g. Ref.~\cite{Ichikawa:2006dt}) to see whether the present Ly$\alpha$ data would hint for non-standard physics.\footnote{
It may suggest that the effective number of neutrinos increases after BBN. In Ref.~\cite{Ichikawa:2007jv}, it is shown that such scenario is feasible by decaying particles.}

\begin{table}[htdp]
\caption{Comparison of $N_\nu$ constraints using various data set combinations. ``All" refers to WMAP3 + other CMB + Ly$\alpha$ + galaxy power spectrum (SDSS main sample + 2dF) + SDSS baryon acoustic oscillation (BAO) + Supernovae Ia (SN). See Ref.~\cite{Seljak:2006bg} for details.}
\begin{center}
\begin{tabular}{|c|c|c|}
\hline
  & 95\% limit & Data set \\
  \hline
Seljak et al. \cite{Seljak:2006bg} & $N_\nu = 5.3^{+2.1}_{-1.7} $ & All \\
 & $N_\nu = 4.8^{+1.6}_{-1.4}$ & All + HST \\
 & $N_\nu = 6.0^{+2.9}_{-2.4}$ & All $-$ BAO \\
 & $N_\nu = 3.9^{+2.1}_{-1.7}$ & All $-$ Ly$\alpha$ \\
 & $N_\nu = 7.8^{+2.3}_{-3.2}$ & WMAP3+SN+SDSS(main) \\
 & $N_\nu = 3.2^{+3.6}_{-2.3}$ & WMAP3+SN+2dF\\
 & $N_\nu = 5.2^{+2.1}_{-1.8}$ & All-2dF-SDSS(main) \\
\hline
\hline
This paper &  $N_\nu = 3.1^{+5.1}_{-2.2}$ & WMAP3+SDSS(LRG) \\
\hline
\end{tabular}
\label{tab:Nnucomparison}
\end{center}
\end{table}

A cosmological constraint on $N_\nu$ can also be obtained from the primordial $^4$He abundance $Y_p$. While $^4$He has logarithmic dependence on the baryon-to-phton ratio $\eta$, the only parameter in the standard BBN, it is very sensitive to $N_\nu$ since it modifies the expansion rate during the BBN period and shifts the epoch of the neutron-to-proton ratio freeze-out. The deuterium, D, constrains these parameters almost in the opposite manner. It is quite sensitive to $\eta$ but has only mild dependence on $N_\nu$. More details are found in e.g. Refs.~\cite{Kneller:2001cd,Barger:2003zg,Cyburt:2004yc}. 
Actually, the BBN bound is more conventional than the structure formation constraint but it has somewhat checkered history since it is very difficult to estimate systematic errors for deriving the primordial abundance from $^4$He observations. 
Although the D abundance is often considered to more robustly probe the primordial abundance, since it does not have much sensitivity on $N_\nu$ as mentioned above, systematic errors for $N_\nu$ estimation are dominated by those of $^4$He.
For example, recent studies have shown the importance of underlying stellar absorption \cite{Olive:2004kq,Fukugita:2006xy}. This leads to significant increase in $Y_p$ and enlarged errors. 
Therefore, the most recent $N_\nu$ constraints $N_\nu = 3.14^{+0.70}_{-0.65}$ (68\% C.L.) \cite{Cyburt:2004yc}  
%
%
has a higher central value and larger errors than the earlier results. Nevertheless, the current BBN bound is significantly tighter than our WMAP+LRG bound and is completely covered by our bound (this is not the case for a so-called MeV reheating scenario. The significance of BBN and CMB/LSS is reversed in this scenario. We comment on it below). At this stage, we can say that the present CMB plus galaxy clustering data provides a complementary constraint to the BBN. Our analysis shows consistency between the constraints derived from totally different physical processes and at distant epochs providing a strong  support for standard cosmology, but relatively large error bars still leave some room for non-standard physics. 

Lastly, let us comment on the implication of our results for MeV-scale
reheating scenarios.
Since the reheating temperature is an important but
not yet known parameter that characterizes the early evolution of the
universe, it is valuable to derive an observational constraint. 
As shown in Sec.~\ref{sec:mevreh}, our estimation of $N_\nu$,
in particular the lower bound of $N_\nu$, provides us with the lower
bound on the reheating temperature as $T_R > 2$ MeV, once we use the
relation between $N_\nu$ and the reheating temperature given in
Ref.~\cite{Ichikawa:2005vw}.
We caution that on the contrary to CMB+LSS bound, the lower bound on $N_\nu$
obtained from $Y_p$ such as the one of Ref.~\cite{Cyburt:2004yc} we
quoted above cannot be taken at face value. This is essentially
because $N_\nu < 3$ in low-reheating scenario implies not only less
radiation density but also less neutron-to-proton conversion rate,
which greatly affect the $^4$He yields by BBN. The latter effect was
not taken into account when deriving the bound on $N_\nu$
in Ref.~\cite{Cyburt:2004yc}. That is why it cannot be applied to the 
MeV-scale reheating scenarios.
It turns out that $Y_p=0.249\pm 0.009$ \cite{Olive:2004kq,Cyburt:2004yc}, 
which does not reject relatively large value of $Y_p$, does not
give meaningful lower bound on $T_R$  (see Fig.~\ref{fig:TR_Nnu_He4} (b) ).
To derive a lower bound on $T_R$ from BBN data alone, one needs a concrete
upper bound on the $^4$He abundance, which is difficult to obtain due to 
a possibly large systematic error.  More detailed discussion is given in
Ref.~\cite{Ichikawa:2005vw}.
At present, CMB+LSS do better job in setting lower bound on $T_R$.
It is quite intriguing that we have obtained the lower bound $T_R > 2$
MeV, which is just before BBN begins, even without resort to the BBN
data. This is because the decoupling of the weak interactions
accidentally occurs immediately before the BBN epoch.  Due to this
coincidence, we were able to derive the concrete and tight bound on
$T_R$.  So far, it has been considered that the observations on the light-element
abundances are indispensable to probe the BBN epoch. However,
our results unambiguously show that one can extract informations on
the universe at $T = O({\rm MeV})$ by CMB+LSS data, and that 
one doesn't have to rely on the observed light-element abundances, which may
have large systematic errors. When the PLANCK data becomes available, 
the lower bound can be improved up to $\sim 5$MeV,  the decoupling 
temperature of the muon and tau neutrinos. 


\end{document}